\documentclass[journal]{IEEEtran}
\usepackage{graphicx}

\begin{document}
\title{Geant4 in Scientific Literature}

\author{Maria Grazia Pia,
	Tullio Basaglia,
        Zane W. Bell,
        Paul V. Dressendorfer% <-this % stops a space
\thanks{Manuscript received November 20, 2009.}% <-this % stops a space
\thanks{T. Basaglia is with CERN, CH-1211 Geneva 23, Switzerland.}%
\thanks{Z. W. Bell is  the Oak Ridge National Laboratory, Oak Ridge TN 37830; 
Oak Ridge National Laboratory is managed by UT-Battelle, LLC,
for the U.S. Department of Energy under contract DE-AC05-00OR22725.}%
\thanks{P. V. Dressendorfer is a consultant and Editor-in-Chief, IEEE TNS. }%
\thanks{M. G. Pia is with INFN Sezione di Genova, 
(telephone: +39 010 3536328, e-mail: MariaGrazia.Pia@ge.infn.it).}%
}

\maketitle
\pagestyle{empty}
\thispagestyle{empty}

\begin{abstract}
The Geant4 reference paper published in Nuclear Instruments and Methods 
A in 2003 has become the
most cited publication in the whole Nuclear Science and Technology category
of Thomson-Reuter's Journal Citation Reports. It is currently the
second most cited article among the publications authored by two
major research institutes, CERN and INFN.
An overview of Geant4 presence (and absence) in scholarly literature
is presented; the patterns of Geant4 citations are quantitatively
examined and discussed. 

\end{abstract}

%\begin{IEEEkeywords}
%IEEEtran, journal, \LaTeX, paper, template.
%\end{IEEEkeywords}

\section{Introduction}
\IEEEPARstart{A}{previous} 
studies \cite{swpub} have highlighted that software-oriented
publications are largely underrepresented in scholarly literature
related to particle physics, with respect to hardware-oriented ones.
Nevertheless, a relatively recent software paper, describing
the Geant4 Monte Carlo system \cite{g4nim}, has
become the most cited publication in the Nuclear Science 
and Technology category defined by Journal Citations Reports \cite{jcr}.

Geant4 is an object oriented toolkit, which provides a 
wide set of tools for the simulation of particle interactions
with matter.
Its development started at the end of 1994 and was motivated
by the requirements of the experiments at the LHC (Large Hadron
Collider) at CERN; nevertheless, since its first release at the end
of 1998, Geant4 has been used by a large community in a variety of 
multi-disciplinary experimental applications beyond its original
scope.

Despite the wide popularity of this software system,
there is limited quantitative documentation of its impact on the
production of physics results, its contribution to technological 
developments, its role in high energy physics and the relative extension 
of its use in this field with respect to other domains.

This paper presents a quantitative analysis of citation patterns
related to Geant4 reference publications \cite{g4nim}, \cite{g4tns}.
Through these data we illustrate the role played by 
Geant4 in experimental physics prior to LHC startup.

\section{Data sources and analysis method}
\label{sec_data}

The main source of data for this study is Thomson-Reuters' Web 
of Science \cite{isiweb}.
The subscription to which the authors had access covers the period since 
1990 to the present date.
Together with publication data, the Web of Science includes a set of 
tools for searching the database and analyzing the search results.
The citing papers were identified through the tools available in the 
Web of Science; the
analysis was restricted to those published before 2009, 
to avoid evolutions of the primary data sample during  the analysis process.

In the course of the analysis, Thomson-Reuters introduced some changes 
in the classification of papers in the Web of Science,
concerning conference proceedings publications, which affected the 
results of various data selections.
The configuration management applied in the analysis process ensured
the reproducibility of consistent results in the course of the
project, despite the changes in the database: the primary data sample
of citations could be reproduced within approximately 1\% throughout 
the duration of the study, and the
outcome of its analysis remained consistent.

According to the latest version of the Thomson-Reuters' database 
used for this study (on October 13, 2009), the
selected data sample consisted of 
1089 papers citing \cite{g4nim} and 127 papers citing \cite{g4tns}.

Complementary analyses were based on publishers' web interfaces
providing full-text search capabilities: the American Physical Society
(APS), Elsevier and IEEE.

Most of the analyses were performed through automated tools provided
by the ISI Web of Science and the publishers; nevertheless, some of
them, requiring more detailed appraisals than the information
available through automated tools, involved a manual inspection of the
publication records.

\section{Monte Carlo in physics and technology literature}
\label{sec_mc}

A preliminary analysis concerned the role played by Monte Carlo
simulation in physics and technological literature pertinent to experimental
particle and nuclear physics, and related research fields such as astronomy
and medical physics.

A set of well known codes was considered for this purpose: 
EGS \cite{egs4,egs5,egsnrc}, FLUKA \cite{fluka,flukachep}, 
GEANT 3 \cite{geant3} and Geant4 \cite{g4nim,g4tns}, 
MCNP \cite{mcnp5,forster,mcnpx} and Penelope \cite{penelope}. 
This selection is not meant to be exhaustive, rather representative 
of the field.

Not all these Monte Carlo systems can be associated with a reference 
publication in an archival journal: for some of them the references are
institutes' reports or contributions to conference proceedings.
Therefore, this analysis was based on the mention of the codes in the 
literature, rather than the citation statistics of proper reference articles.

The results were collected by means of the full-text tools
provided by a few major publishers in the field through their web interfaces.
%the American Physical Society, Elsevier and IEEE.
The search pattern identified the various versions of the codes and
naming variants commonly mentioned in the scientific literature 
(e.g. MCNP and MCNPX, EGS version 4 and 5 etc.).
Papers mentioning FLUKA were manually inspected to ascertain whether
they concerned the standalone FLUKA code or the FLUKA package interfaced
to GEANT 3: in the latter case, they were associated with GEANT 3.

The journals examined in this analysis were the Physical Reviews (A,
B, C, D, E and Letters) as representative of physics journals, IEEE
Transactions on Nuclear Science (TNS) and Nuclear Instruments and
Methods (NIM), both A and B, as representative of technology journals.
Among the Physical Reviews, most of articles mentioning the Monte Carlo codes
considered in this study are published in Physical Review Letters, 
Physical Review D and C (97\% of them over the years from 1990 through 2008);
therefore the analysis focused on these three journals.

\begin{figure}
\centering
\includegraphics[angle=0,width=9cm]{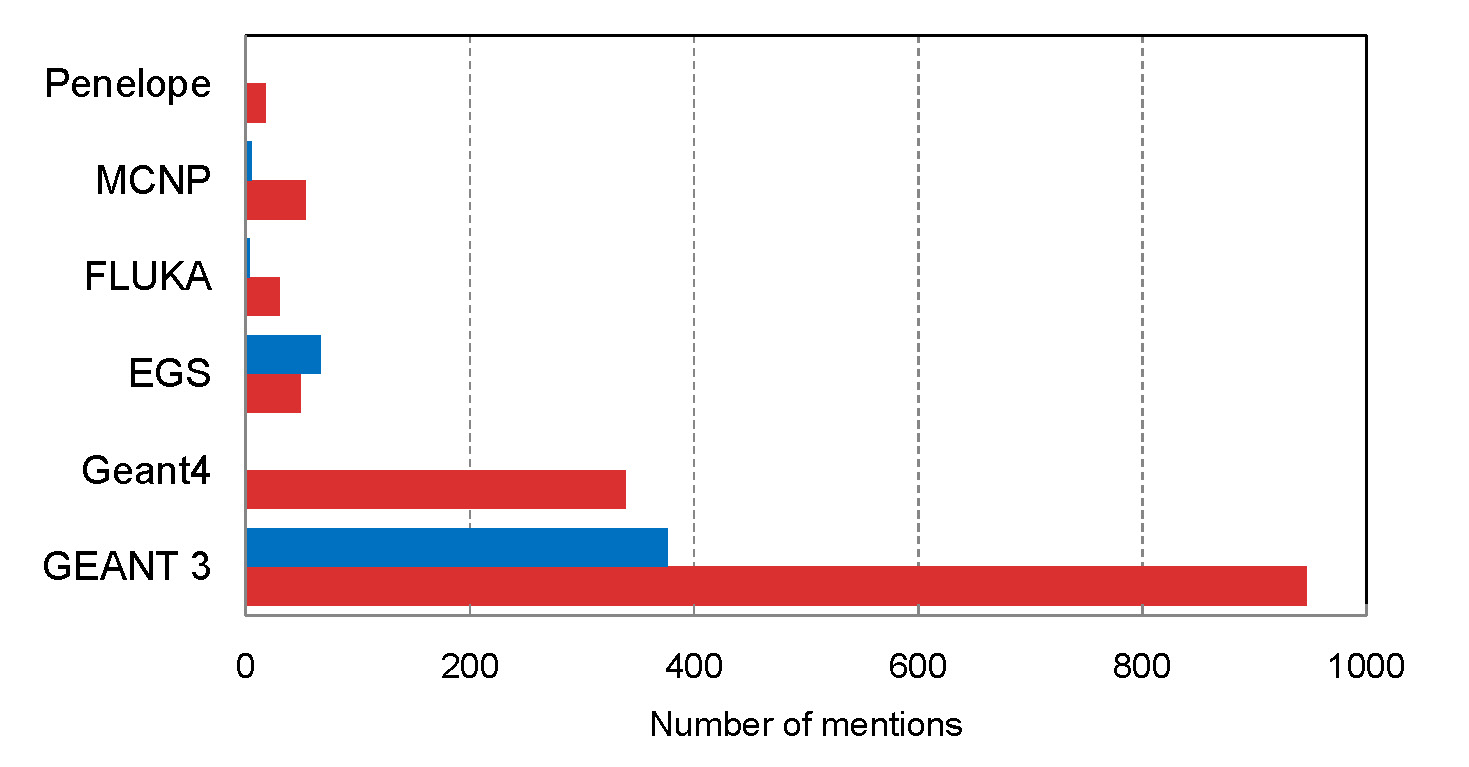} 
\caption{Number of papers mentioning well known Monte Carlo codes in APS
journals Physical Review C, D and Physical Review Letters, 
over the period 1990-1999 (blue) and 2000-2008 (red).}
\label{fig_mc_aps_1990_2008}
\end{figure}

\begin{figure}
\centering
\includegraphics[angle=0,width=9cm]{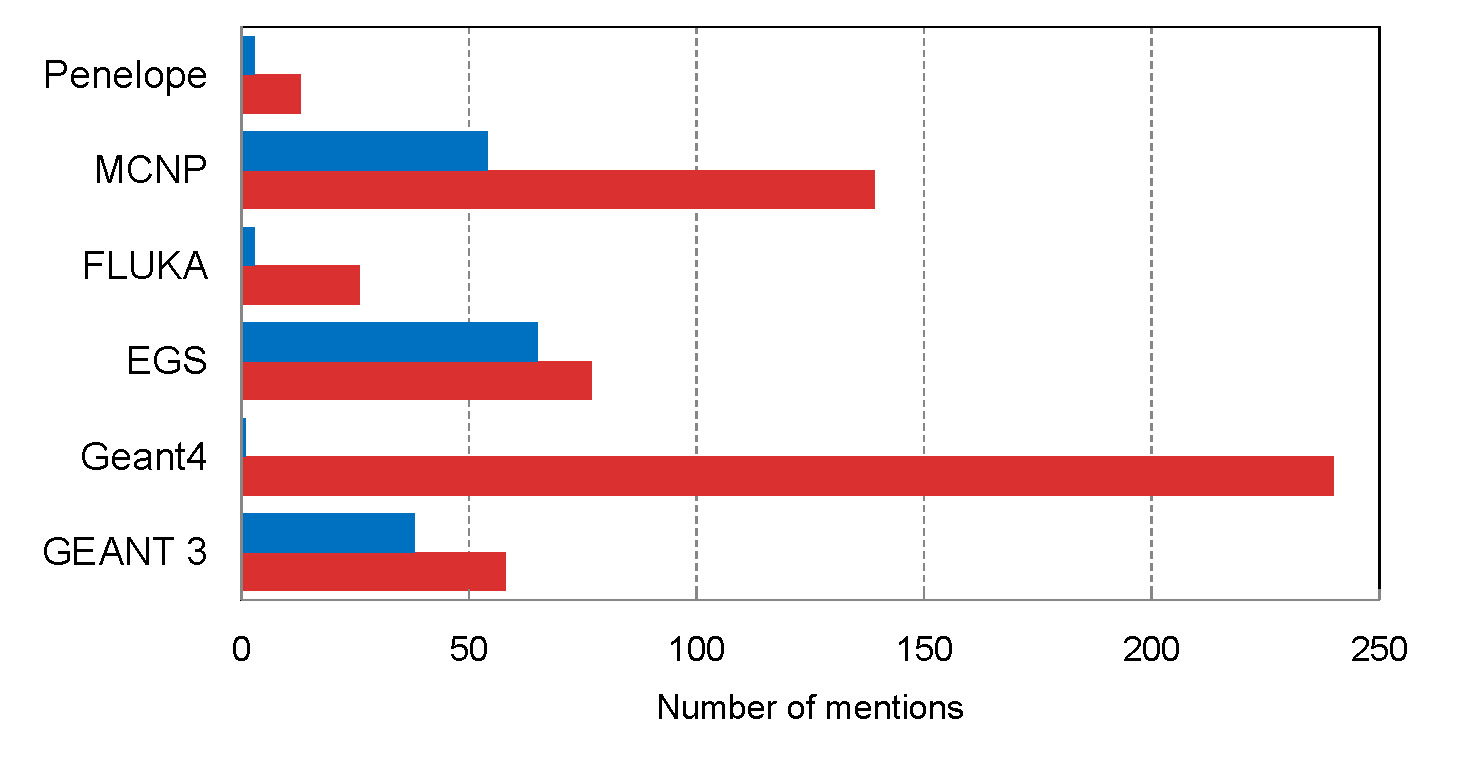} 
\caption{Number of papers mentioning well known Monte Carlo codes in 
IEEE Transactions on Nuclear Science (TNS) over the period 1990-1999 (blue)
and 2000-2008 (red).}
\label{fig_mc_tns_1990_2008}
\end{figure}

\begin{figure}
\centering
\includegraphics[angle=0,width=9cm]{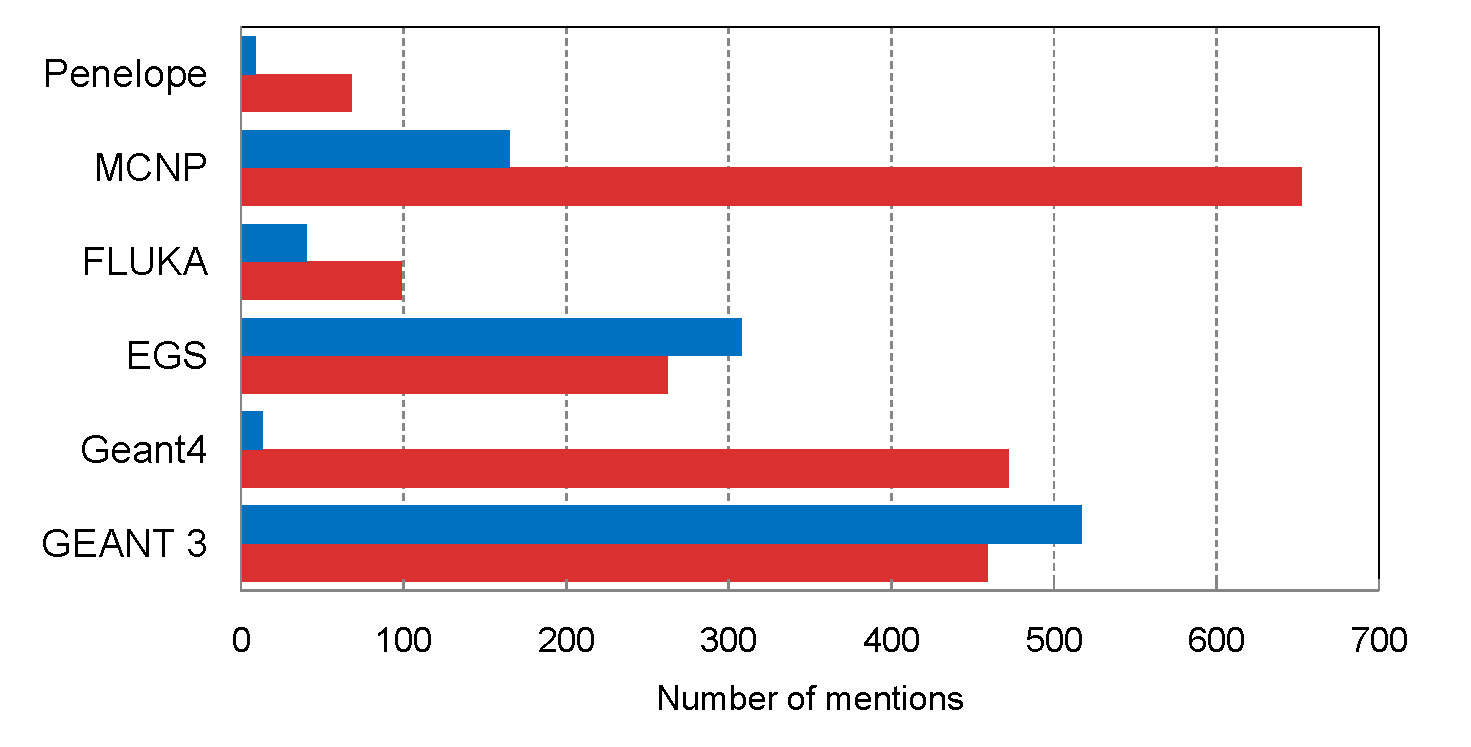} 
\caption{Number of papers mentioning well known Monte Carlo codes in 
Nuclear Instruments and Methods (NIM) A and B over the period 1990-1999 (blue)
and 2000-2008 (red).}
\label{fig_mc_nim_1990_2008}
\end{figure}

One can observe in Figs. \ref{fig_mc_aps_1990_2008}-\ref{fig_mc_nim_1990_2008}
that, as a general trend, the use of Monte Carlo codes has increased both 
in physics and technology journals.

Monte Carlo simulation enables 
the production of physics results and supports technological research.
Out of the 13407 papers  published by NIM and the 2630 ones published by TNS 
over the 2004-2008 period, respectively 45\% and 58\% mention ``simulation''
or ``Monte Carlo'' in the text.
This pattern appears correlated with modeling: over the same period,
64\% of the articles published in TNS mention ``model'' or ``modeling'',
out of which 44\% also mention ``simulation'' or ``Monte Carlo''.

\begin{figure}
\centering
\includegraphics[angle=0,width=9cm]{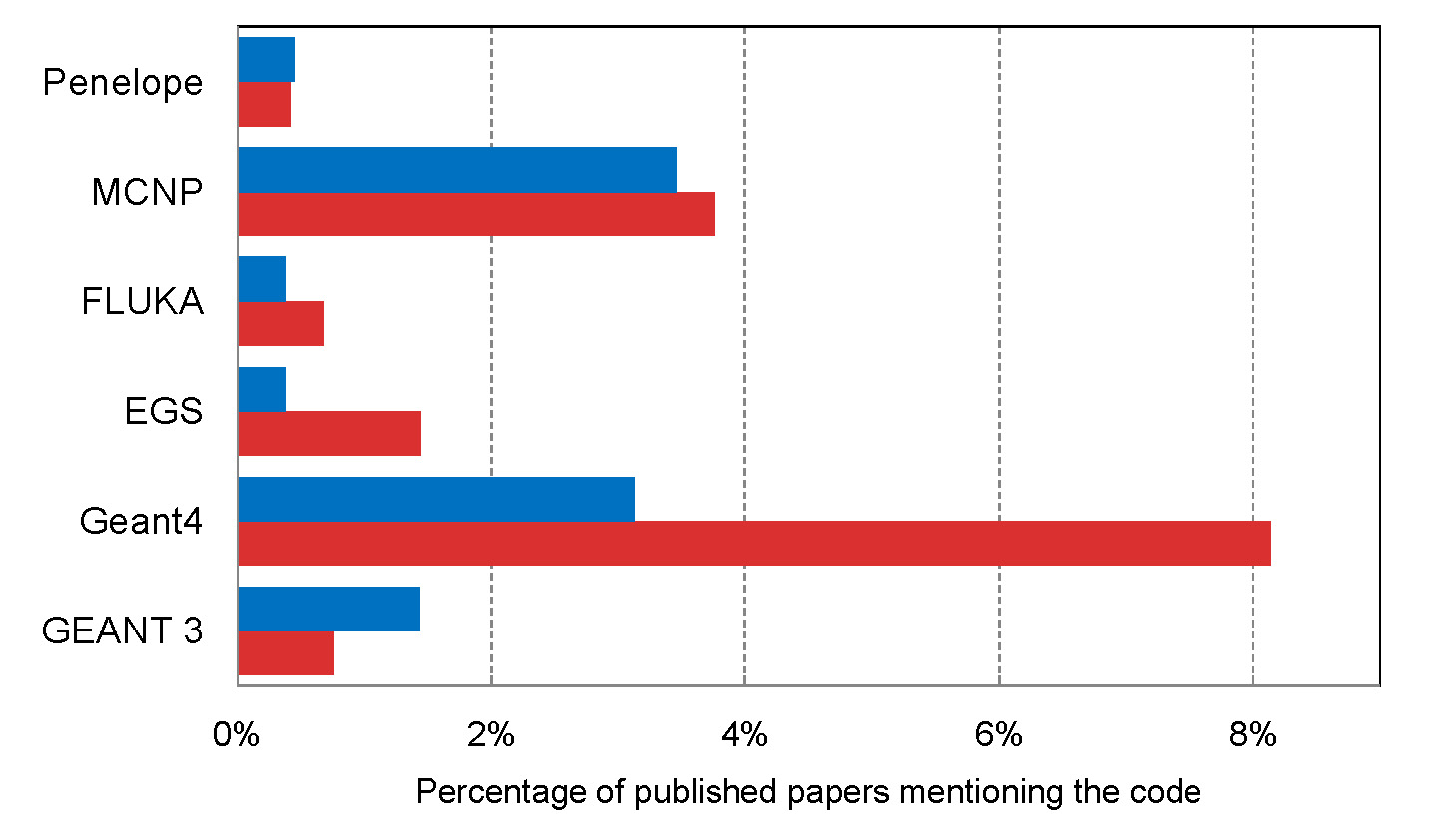} 
\caption{Fraction of papers published in 
2004-2008 mentioning well known Monte Carlo codes:  
NIM A and B (blue) and TNS (red).}
\label{fig_mc_nim_tns}
\end{figure}

The papers mentioning the considered Monte Carlo codes amount to 15\%
and 9\% of those published respectively by TNS and NIM in 2004-2008;
their distributions are shown in Fig. \ref{fig_mc_nim_tns}.
The discrepancy of these values with respect to the fraction of
articles mentioning ``Monte Carlo'' or ``simulation'' suggests that a
significant portion of Monte Carlo simulation in the field covered by
these journals involves other codes.

\begin{figure}
\centering
\includegraphics[angle=0,width=9cm]{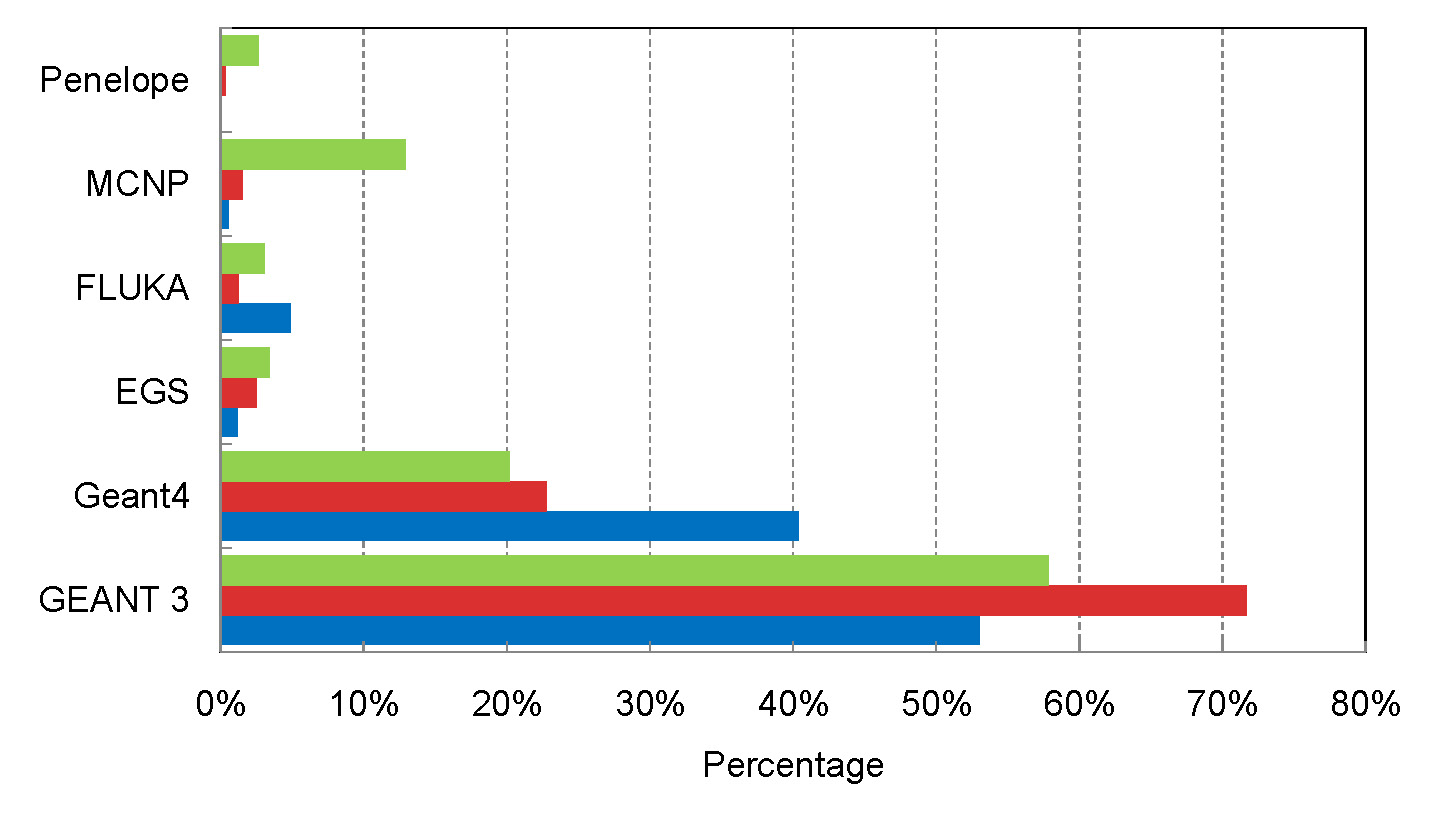} 
\caption{Fraction of papers published in 
2004-2008 mentioning well known Monte Carlo codes:  
Physical Review D (blue), Physical Review Letters (red) and Physical Review C
(green).}
\label{fig_mc_aps}
\end{figure}

The papers published in Physical Review C, D and Letters, which mention 
well known Monte Carlo codes, amount to 933 over the 2004-2008 period; 
their distribution is shown in Fig. \ref{fig_mc_aps}.

An evident result that emerges from this analysis is the continuing wide use of
GEANT 3 in recent years, despite the fact that the latest version of this code
(3.21) dates back to 1994.
The use of GEANT 3 is more extensive in physics journals over
technology ones, and in NIM over TNS.

Various factors contribute to the continuing significant presence of GEANT 3 
in fundamental physics journals.
A large number of physics publications in APS journals derive from 
experiments that started taking data before Geant4 was first released, 
or shortly after its release, when this code was not established yet, 
and are still actively analyzing their data to produce physics results.
Despite the more advanced features offered by Geant4 with respect to GEANT 3,
moving to a new code would represent a major risk and effort for a mature
experiment in the course of its physics analysis, and could affect the
systematics of the results.
For this reason, most of these experiments 
tend to maintain a consistent simulation production environment over
their lifecycle, and still rely on the Monte Carlo system, 
GEANT 3, on which they initially based their simulations.
The requirement for many experiments to keep their simulation environment
unchanged throughout their lifecycle is confirmed by the fact that, among the
papers published between 2004 and 2008, some mention older GEANT versions
than the latest 3.21 release, extending down to version 3.13.

In other cases of more recent 
high energy and nuclear physics experiments, the use 
of GEANT 3 is motivated by the
decision of pursuing the experimental activity in a procedural 
programming environment, thus avoiding the transition to the object oriented 
technology associated with Geant4, which is perceived as a demanding 
investment of resources.

The type of research within the scope of technology journals is more likely
to profit from the new functionality and modern software technology offered by
Geant4; in this respect, TNS appears more open than NIM towards the use of
Geant4 over GEANT 3.

With the exception to some extent of MCNP, which is relatively often
mentioned in Physical Review C, GEANT 3 and Geant4 jointly are by far the 
most widely used simulation environment in fundamental particle and nuclear 
physics experiments,by far outstripping EGS, FLUKA and Penelope.
Among technology journals, MCNP plays a significant role jointly with 
Geant4 and GEANT 3.

\section{Geant4 citation patterns}

\begin{figure}
\centering
\includegraphics[angle=0,width=9cm]{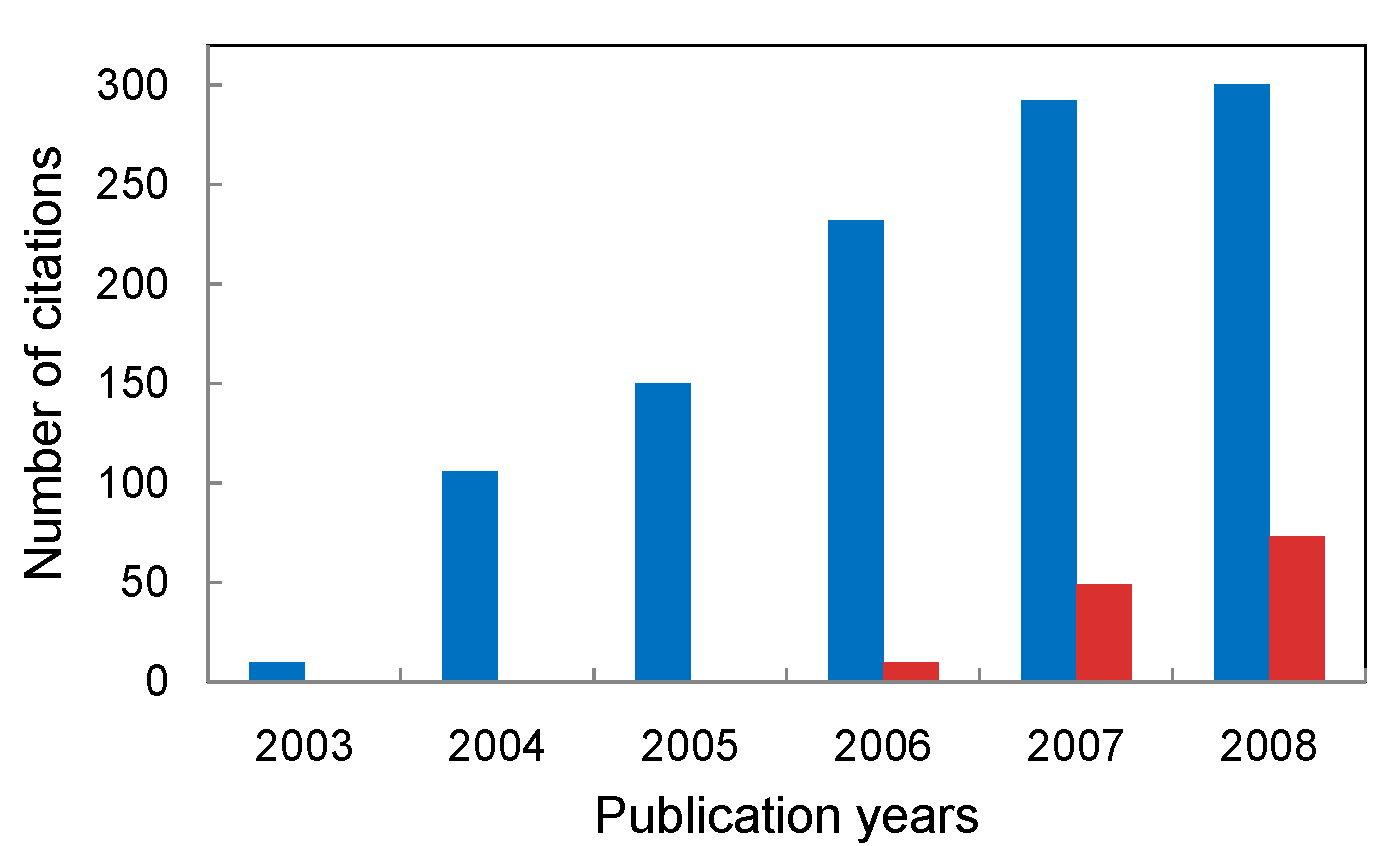} 
\caption{Number of citations per year collected by \cite{g4nim} (blue)
and \cite{g4tns} (red). }
\label{fig_years}
\end{figure}

The distribution of Geant4 citations since the publication 
of \cite{g4nim} is shown in Fig. \ref{fig_years}.
One observes a growing trend as a function of time, that seems to be
slower in recent years;
however, this effect should be verified over a more extended time scale,
and could be affected by major events in experimental research, such as
the start of LHC operation foreseen at the end of 2009. 
Currently, \cite{g4nim} averages approximately a citation per day.

A striking feature of Fig. \ref{fig_years} is the much smaller number 
of citations collected by 
\cite{g4tns} with respect to \cite{g4nim}, despite the fact that both 
publications are indicated on the Geant4 web page as references for the code.
Nevertheless, citations to  \cite{g4tns} contribute approximately 10\% 
to the 2006 portion of TNS 2008 impact factor.
The correct citation of this reference at the same level as \cite{g4nim}
would have contributed to raise the journal's impact factor considerably
in 2007 and 2008.

Due to the limited statistical significance of the citation sample
associated with \cite{g4tns}, the analysis of citation patterns has been
focused on \cite{g4nim}.
The results related to \cite{g4tns} confirmed in general the 
citations patterns observed with \cite{g4nim}; the few relative
differences are discussed in the following sections.

Reference \cite{g4nim} is the second most cited paper for CERN and INFN
(over the period covered by the ISI Web of knowedge acessible to the authors);
it was ranked fourth relative to both institutes at the time of publication 
of \cite{swpub}.

A large fraction (approximately 17 \%) of the citations to
\cite{g4nim} derive from a single high energy physics experiment,
BaBar \cite{babar}, located at Stanford Linear Accelerator Center;
this feature could affect the overall appraisal of the citation
patterns.
Some of the analyses have been performed not only on the whole citation data
sample, but also on a subset excluding BaBar contributions, to evaluate the
possible bias introduced by the large weight of this experiment in the 
overall picture. 

\subsection{Geographical distribution}

The distributions of citations to \cite{g4nim} by geographical
area and country are shown in Tables
\ref{tab_continent} and \ref{tab_country}. The latter is 
limited to the countries
ranked in the top 10 positions in terms of number of citations;
the equivalent distribution 
excluding BaBar papers is  in  Table \ref{tab_country_nobabar}.
The totals in these tables and in the following  ones 
(\ref{tab_inst}-\ref{tab_inst_nocern}) are larger than 100\%, 
since the co-authors of a
paper may come from multiple geographical areas, countries and institutes.
 
The largest number of citations come from Europe as a geographical
area and from the United States as an individual country.
The role of the USA as the country contributing the largest number of citations
is more evident in the citation sample excluding BaBar; however, the 
major features of the citing country distribution are similar over the 
two data samples, apart from the more prominent position of Japan in the 
sample excluding BaBar.

%\begin{figure}
%\centering
%\includegraphics[angle=0,width=9cm]{continent.jpg} 
%\caption{Geographical areas of the citations to \cite{g4nim}.}
%\label{fig_continent}
%\end{figure}

\begin{table}
\begin{center}
\caption{Geographical areas of the citations to \cite{g4nim}.}
\label{tab_continent}
\begin{tabular}{|l|c|}
\hline 
\textbf{Geographical Area}		&\textbf{Percentage (\%)}      \\
\hline
Europe					& 69 \\ 
North America				& 49 \\
Asia					& 31 \\
Russia + former Soviet Union countries	& 27 \\
South America				& 2.4 \\
Oceania 				& 2.4\\
Africa 					& 0.9\\
\hline
\end{tabular}
\end{center}
\end{table}

\begin{table}
\begin{center}
\caption{Geographical origin of the citations to \cite{g4nim}: 
top 10 countries.}
\label{tab_country}
\begin{tabular}{|l|c|}
\hline 
\textbf{Country}		&\textbf{Citations (\%)} \\
\hline 
USA		&47 \\
Germany		&32 \\
Italy		&30 \\
France		&29 \\
England		&28 \\
Russia		&26 \\
Spain		&25 \\
Canada		&22 \\
Netherlands	&21 \\
Scotland	&19 \\
\hline     
\end{tabular}
\end{center}
\end{table}

\begin{table}
\begin{center}
\caption{Geographical origin of the citations to \cite{g4nim}: 
top 10 countries, escluding citations associated with the BaBar experiment.}
\label{tab_country_nobabar}
\begin{tabular}{|l|c|}
\hline 
\textbf{Country}		&\textbf{Citations (\%)} \\
\hline 
USA         &35 \\
Germany     &18 \\
Italy       &16 \\
Switzerland &15 \\
France      &15 \\
England     &14 \\
Japan       &13 \\
Russia      &11 \\
Spain       &10 \\
Canada      &6  \\
\hline     
\end{tabular}
\end{center}
\end{table}

The distribution of institutes citing \cite{g4nim} is strongly biased by 
BaBar citations.
It is led by INFN,
followed by a long list of more than 60 institutes contributing 
approximately the same number of citations, almost entirely associated 
with BaBar. The first 10 institutes in the list, according to the
order defined by the Web of Science, are reported in Table \ref{tab_inst};
the statistics are based on a total of 1070 citing articles.

Excluding BaBar papers from the analysis, the 10 institutes 
contributing the largest number of citations are reported in 
Table \ref{tab_inst_nobabar}. 
The list is still led by INFN, followed by CERN;
two Japanese institutes appear in the top ranks.
These statistics are based on 882 citing articles.

\begin{table}
\begin{center}
\caption{Origin of the citations to \cite{g4nim}: 
top 10 institutes.}
\label{tab_inst}
\begin{tabular}{|l|c|}
\hline 
\textbf{Institute}		&\textbf{Citations} \\
\hline 
INFN                     &288 \\
Rutherford Appleton Lab. &207 \\
Univ. Milan              &205 \\
Univ. Liverpool          &203 \\
Univ. Valencia           &203 \\
Harvard Univ.            &199 \\
Univ. Padua              &199 \\
Univ. Roma La Sapienza   &199 \\
Univ. Calif Los Angeles  &198 \\
Ohio State Univ.         &196 \\
\hline     
\end{tabular}
\end{center}
\end{table}

\begin{table}
\begin{center}
\caption{Origin of the citations to \cite{g4nim}: 
top 10 institutes, excluding references associated with the 
BaBar experiment.}
\label{tab_inst_nobabar}
\begin{tabular}{|l|c|}
\hline 
\textbf{Institute}		&\textbf{Citations} \\
\hline 
INFN                    &105 \\
CERN                    &82  \\
Univ. Tokyo             &42  \\
Univ. Valencia          &37  \\
Kyoto Univ.             &34  \\
Russian Acad. Sci.      &28  \\
JINR                    &26  \\
Univ. Oxford            &26  \\
Univ. Sheffield         &26  \\
Rutherford Appleton Lab &25  \\
\hline     
\end{tabular}
\end{center}
\end{table}

\begin{table}
\begin{center}
\caption{Origin of the citations to \cite{g4nim}: 
top 10 institutes, excluding references associated with the 
BaBar experiment and CERN authors.}
\label{tab_inst_nocern}
\begin{tabular}{|l|c|}
\hline 
\textbf{Institute}		&\textbf{Citations} \\
\hline 
INFN              &70 \\
Univ. Tokyo       &39 \\
Kyoto Univ.       &28 \\
RIKEN             &21 \\
Univ. Valencia    &21 \\
Vanderbilt Univ.  &21 \\
NASA              &20 \\
Univ. Liverpool   &20 \\
Univ. Michigan    &19 \\
Harvard Univ.     &18 \\
\hline     
\end{tabular}
\end{center}
\end{table}

CERN plays a major role in relation to Geant4; Geant4's development itself
was motivated by the requirements of the experiments at the LHC, 
and the Geant4 release infrastructure is hosted by CERN.
Once BaBar citations had been discarded, a further selection was performed 
to evaluate the degree of correlation with CERN by excluding the papers 
involving authors with CERN affiliations; the resulting data sample
included 800 papers.
The 10 institutes contributing
the largest number of citations in the selected sample are listed in 
Table \ref{tab_inst_nocern}: INFN confirms its leading role, followed by
a significant presence of Japanese institutes.

\subsection{Research areas}

\begin{figure}
\centering
\includegraphics[angle=0,width=9cm]{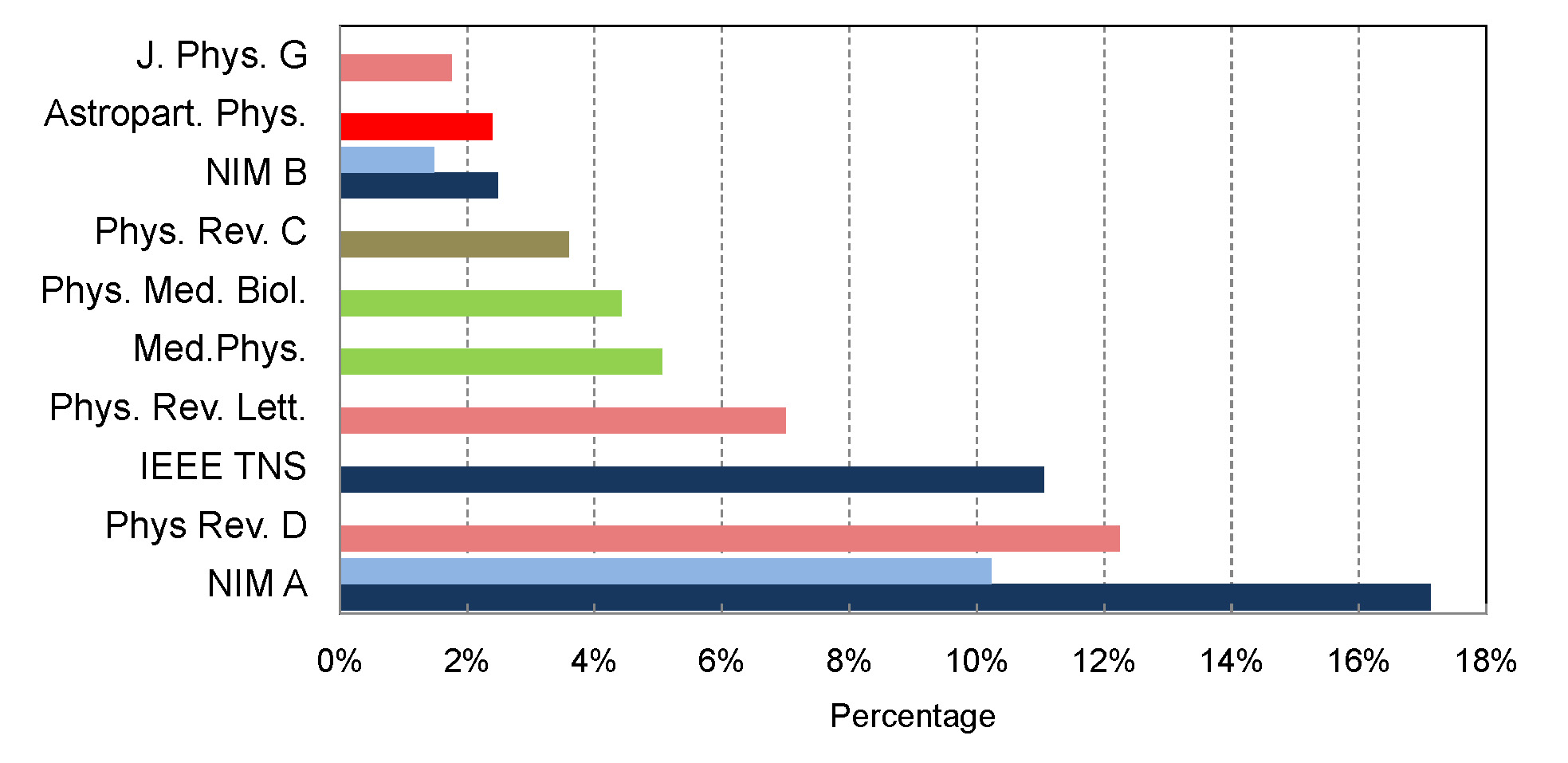} 
\caption{The 10 journals associated with the largest number of citations to 
\cite{g4nim}; the light blue bins show the fraction of NIM A and B citations 
not deriving from conference proceedings. }
\label{fig_journals}
\end{figure}

The journals providing citations to \cite{g4nim} encompass a widely
multi-disciplinary scope; they include particle and nuclear physics, 
technology,
astrophysics and medical physics journals, and fields as diverse as geophysics,
plasma science and materials science.
The top 10 are shown in Fig. \ref{fig_journals}; 
the statistics are based on 1086 citing papers.
Regarding NIM, both the total number of citing articles, and the number 
resulting from the exclusion of conference proceedings papers are shown;
in the latter case no further renormalization was performed to 
account for the modified citation sample size in the calculation of 
the fraction of papers relative to the other journals.
Multi-disciplinary technology journals appear to be the major source
of citations, together with HEP publications in Physical Review D and 
Physical Review Letters.

The references to \cite{g4tns} exhibit a different pattern: while for 
\cite{g4nim}  technology and
physics journals are the major sources of citations, 
medical physics journals, together with TNS and NIM, originate most of the
citations to  \cite{g4tns}.
However, due to the limited statistical significance of the citation
sample associated with \cite{g4tns}, it is hard to derive any firm
conclusions from this observation.

An effort was invested to identify the research areas from which the citations
to \cite{g4nim} derive, and to estimate their relative 
contribution quantitatively.
In some cases the identification was straighforward: for instance, papers
published in journals characterized by well-defined scope (e.g. Medical 
Physics, Physical Review D etc.) were attributed to the related research
domain.
Other criteria involved the association with experiments, projects and 
research groups, whose scope of activity is well known in the community.
The papers which could not be attributed to a research area by means of 
automated criteria were inspected manually by examining the 
abstract and, in a few cases, the whole article.
This analysis involved some degree of subjectivity; nevertheless,
we do not think that it introduced any significant bias in the results.
The amount of noise and the incompleteness of the data samples 
deriving from automated searches 
affect the conclusions of the various analyses; the uncertainties of
the results as determined from manual inspection are smaller than 5\%.

\begin{figure}
\centering
\includegraphics[angle=0,width=9cm]{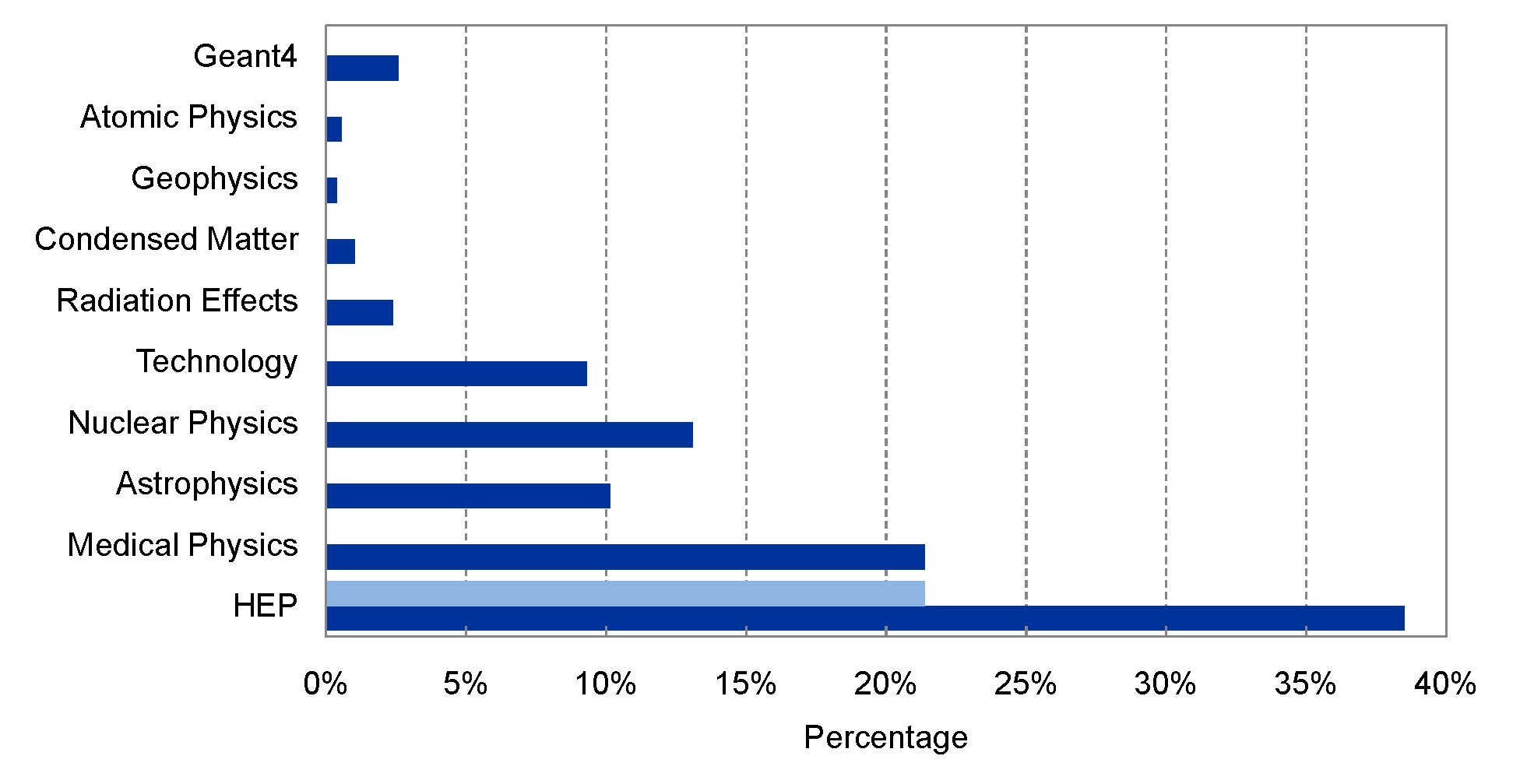} 
\caption{Distribution of citations to \cite{g4nim} across the various 
disciplines; the light blue bin shows the fraction of HEP citations not 
associated with BaBar; the Geant4 bin shows the fraction of citations 
deriving from 
papers by Geant4 developers concerning Geant4 developments. }
\label{fig_field}
\end{figure}

The distribution of research areas contributing citations to  \cite{g4nim}
is shown in Fig. \ref{fig_field}; it is based on the sample of 1086 papers.
High energy physics appears the major source of citations; nevertheless,
if BaBar papers are excluded, the contribution from medical physics becomes
comparable to the one from the rest of HEP.
This result confirms the observation in \cite{swpub} that, while
Geant4 development was originally motivated by high energy physics
requirements and many of its developers are affiliated with high
energy physics laboratories and institutes, Geant4 use extends far
beyond high energy physics; the present analysis provides the first
quantitative estimate of Geant4 application to different 
scientific research areas.

\begin{figure}
\centering
\includegraphics[angle=0,width=9cm]{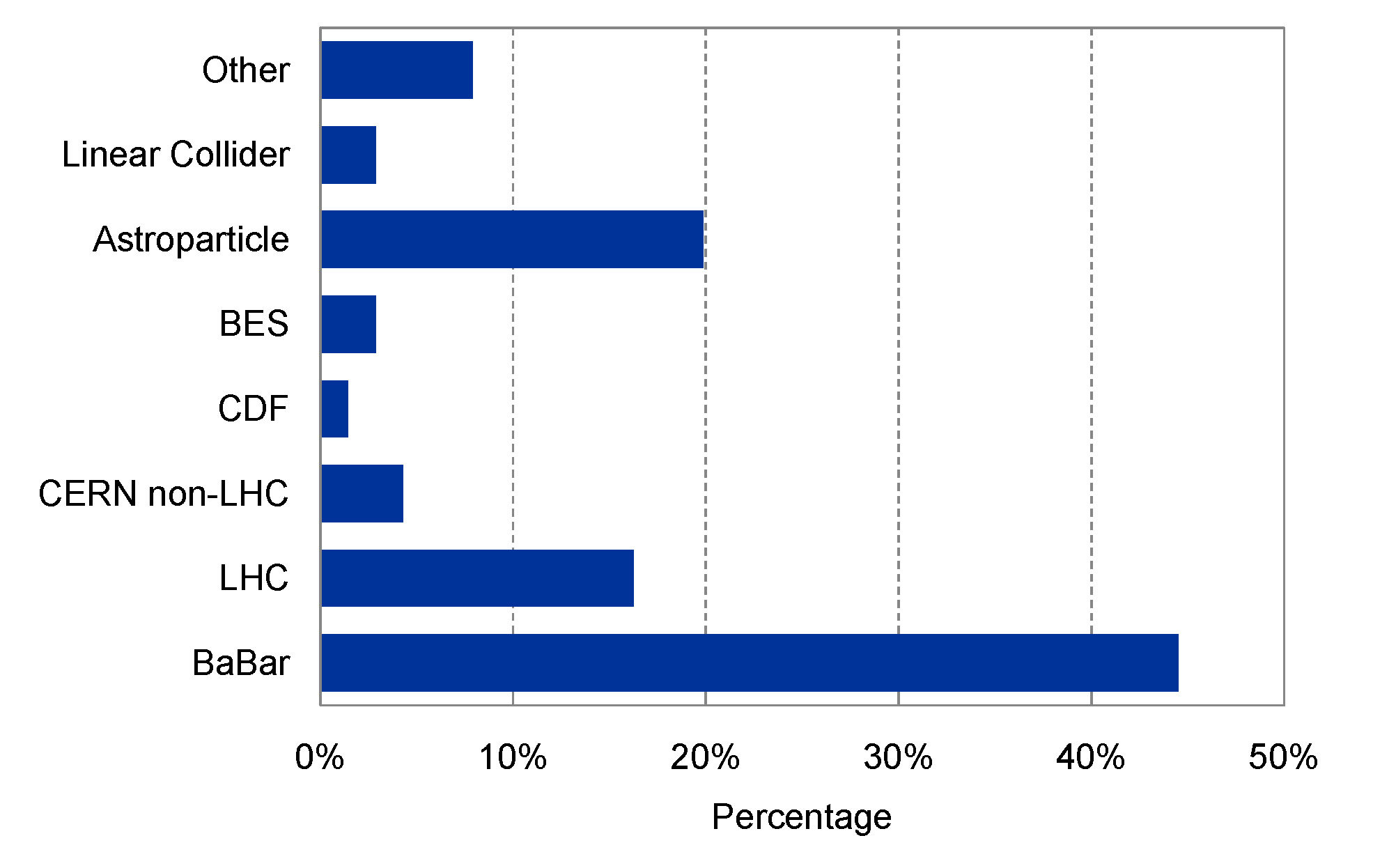} 
\caption{Distribution of citations to \cite{g4nim} across HEP domains. }
\label{fig_hep}
\end{figure}

The 418 papers associated with HEP research were further classified 
according to their pertinent experimental sub-domain. 
The analysis involved manual inspection of the publication records
(abstract and full paper) in the cases where automated criteria could not
identify the proper attribution of a paper.
The results are shown in Fig. \ref{fig_hep}.

By far, the largest number of HEP citations are associated with BaBar.
It is worthwhile remarking that 59\% of the physics papers published by BaBar
over 2004-2008 cite Geant4; this observation confirms the strategic role
played by Monte Carlo simulation, and Geant4 in particular, in the physics
analysis of HEP experiments.
The indications coming from BaBar can be extrapolated as similar expectations
for the LHC experiments, which have based their simulations on Geant4.

The second largest source of HEP citations is astroparticle physics; somewhat
surprisingly, at this stage 
the citations from this field outnumber those related to LHC,
whose experimental program motivated the development of Geant4.
However, these results should be revisited after LHC becomes operational
and the LHC experiments start publishing physics results.

\section{Missing citations}

The Geant4 citation patterns in Nuclear Instruments and Methods A and
IEEE Transactions on Nuclear Science reported in \cite{swpub} showed
that Geant4 is not properly cited in many cases.
This analysis was extended to the larger data sample now available in these 
technology journals and to two additional data samples:
a set of physics journals (the Physical Reviews, published by the 
American Physical Society) and the multi-disciplinary collection of journals
published by Elsevier.

The analysis concerned papers published in 2004-2008, 
where Geant4 is mentioned in text; it verified whether they properly cite
\cite{g4nim}.
The results concerning the fraction of proper citations in 
NIM, TNS and the relevant subset of Physical Reviews
are summarized in Table \ref{tab_missing}.
The papers published in physics journals appear more diligent
at properly citing Geant4 reference than those in technology journals.
However, with respect to the data reported in \cite{swpub}, the fraction of
properly citing papers in NIM has significantly increased, while it
has remained approximately constant in TNS.

A similar analysis over the whole Elsevier journals collection found that
40\% of the publications in these journals correctly cite \cite{g4nim}, 
when Geant4 is mentioned in the text. 

The more recent Geant4 reference \cite{g4tns} appears to be seldom cited:
only 27\% of TNS articles and 10\% of NIM ones published in 2007-2008
properly cite it, when Geant4 is mentioned in the text.
Hardly any paper cites it in fundamental physics journals.

\begin{table}
\begin{center}
\caption{Fraction of papers properly citing \cite{g4nim}}
\label{tab_missing}
\begin{tabular}{|l|c|}
\hline 
\textbf{Journal}	&\textbf{Percentage (\%)}      \\
\hline
NIM A and B		& 51 \\
\hline
TNS 			& 59 \\ 
\hline
Phys. Rev. C		& 64 \\
\hline
Phys. Rev. Lett.	& 93 \\
\hline
Phys. Rev. D 		& 81 \\
\hline
\end{tabular}
\end{center}
\end{table}

\section{Conclusion}

This study documented a detailed, quantitative analysis of 
citation patterns related to Geant4.
It highlighted the major role played by Monte Carlo simulation both in
fundamental nuclear and particle physics research, and in related
technological research; this role has become more visible 
when considering the years 2000-2008 as compared with the years 1990-1999.
%in the recent years with respect to the last decade of the previous century.

Geant4's first reference paper has rapidly become the most cited publication
in Nuclear Science and Technology.
Nevertheless, the use of GEANT 3 remains widespread in particle and nuclear
physics experiments, and is documented in recent publications of physics 
results.

The Geant4 user community is largely multi-disciplinary,
with high energy physics and
medical physics contributing the largest numbers of citations.
Within HEP, the BaBar experiment and the astroparticle community have 
published the largest number of papers citing Geant4. 
However, this paper
reflects the citation sample as in October 2009, a few weeks before
the beginning of LHC commissioning phase;
the results concerning HEP may be subject to change, when LHC starts operating.

Along with the large number of citations collected by Geant4, a large number of
publications do not cite references for the code they use.
There are a number of reasons for this pattern, including authors 
considering Geant4 to be a public domain facility, a perception that 
it is not the result of scientific research, or the fact that many 
well-known Monte Carlo codes lack an associated reference publication 
in a peer-reviewed journal.
%The reasons for this pattern could be multiple, like considering Geant4 as
%a public domain facility, or not perceiving it as the result of scientific 
%research, or the fact that many well-known Monte Carlo codes lack an
%associated reference publication in a scholarly journal.

%this pattern is a source of concern, as it hints that 
%Geant4 - and, more in general, other major Monte Carlo systems - may not be 
%widely perceived by the experimental community as the result of scientific 
%research.

\section*{Acknowledgment}

The authors are grateful to Simone Giani and Jens Vigen for valuable 
discussions.

\end{document}